\documentstyle[12pt,epsfig]{article}
\input epsf

\newcommand{\be}{\begin{eqnarray}}
\newcommand{\ee}{\end{eqnarray}}

\newcommand{\ba}{\begin{array}}
\newcommand{\ea}{\end{array}}

\begin{document}
\rightline{RUB-TP2-17/99}

\begin{center}
{\Large Polarized antiquark distributions from chiral quark-soliton model:
summary of the results}\\[0.5cm]

K.~Goeke$^a$, P.V. Pobylitsa$^{a,b}$,
M.V.~Polyakov$^{a,b}$ and D.~Urbano$^{a,c}$\\[0.3cm]

\footnotesize\it $^a$
Institute for Theoretical Physics II,
Ruhr University Bochum, Germany\\
\footnotesize\it $^b$
Petersburg Nuclear Physics Institute, Gatchina,
St. Petersburg 188350, Russia\\
\footnotesize\it $^c$
Faculdade de Engenharia da Universidade do Porto, 4000 Porto, Portugal

\end{center}
\begin{abstract}
In these short notes we present a parametrization of the results obtained
in the  chiral quark-soliton model
for polarized
antiquark distributions $\Delta\bar u$, $\Delta\bar d$ and
$\Delta\bar s$ at a low normalization point around $\mu=0.6$~GeV.
\end{abstract}

The aim of these short notes is to summarize the results
for the polarized
antiquark distributions $\Delta\bar u$, $\Delta\bar d$ and
$\Delta\bar s$ obtained in refs.~\cite{DPPPW96,PPG99,GPPPSU99}
in the framework of the chiral quark-soliton model.

The chiral quark-soliton model \cite{DPP}
is a low-energy field theoretical  model of the
nucleon structure  which allows a consistent calculations of leading
twist quark and antiquark distributions \cite{DPPPW96}. Due to its field
theoretical nature the quark and antiquark distributions obtained in this
model satisfy all general QCD requirements: positivity, sum rules,
inequalities, etc.

A remarkable prediction of the chiral quark soliton model,
noted first in ref.~\cite{DPPPW96}, is
the strong flavour asymmetry of polarized antiquarks, the  feature
which is missing in other models like, for instance, pion cloud models
(for discussion of this point see Ref.~\cite{SIDIS}).

The fits below are based on the calculations
of Refs.~\cite{DPPPW96,PPG99,GPPPSU99},
generalized to the case of three flavours.
The results of these calculations are fitted by the form
inspired by quark counting rules discussed in Ref.~\cite{Brod}:

\be
\Delta\bar q(x)&=&\frac{1}{x^{\alpha_q}}\
\biggl[A_q (1-x)^5+B_q (1-x)^6\biggr]\,,
\label{fit}
\ee
which leads to
\be
\nonumber
\alpha_u&=&0.0542,\ \alpha_d=0.0343,\ \alpha_s=0.0169 \\
\nonumber
     A_u&=&0.319,\       A_d=-0.185,\      A_s=-0.0366\\
     B_u&=&0.589,\       B_d=-0.672,\      B_s=-0.316\, . \label{eq:param}
\ee
In Fig.~1 we plot the resulting distribution functions.
We note that these functions, obtained in the framework of the chiral quark
soliton model,  refer to the normalization
point of about $\mu=0.6$~GeV.

A few comments are in order here:
\begin{itemize}
\item
 The model calculations are not justified at $x$ close to zero and one.
Therefore the small $x$ and $x\to 1$ behaviours obtained in the
the fit above should be consider as an educated guess only, not
as model prediction.
\item
We estimate that the theoretical errors related to the approximations
($1/N_c$ corrections, $m_s$ corrections, etc.)
done in the model
calculations are at the level of 20\%-30\% for
$\Delta\bar u$ and $\Delta\bar d$, and around 50\% for
$\Delta\bar s$.
The  value of the normalization point is not known exactly, the most
favoured value is $\mu=0.6$~GeV.
\end{itemize}
The measurements of flavour asymmetry of polarized antiquarks, say,
in semi-inclusive DIS \cite{SIDIS}
or in Drell-Yan reactions with polarized protons \cite{DY}
would allow to discriminate between different pictures of the nucleon.

\begin{figure}[tbp]
\epsfxsize=6cm 
\centerline{\epsffile{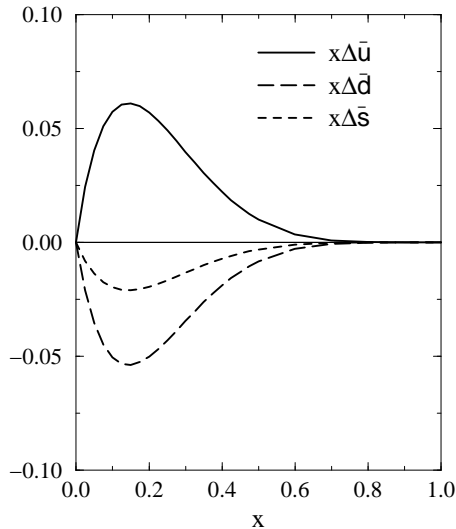}}
\caption{Results for $x\Delta\bar u(x)$, $x\Delta\bar d(x)$ and
$x\Delta\bar s(x)$ at low normalization point obtained in
chiral quark soliton model}
\label{iaa}
\end{figure}

\end{document}